\documentclass[11pt, twoside, a4paper]{article}
\usepackage[english]{babel}
\usepackage{amsmath,amssymb,amsthm,epsfig,color,graphicx}
\usepackage[latin1]{inputenc}
\usepackage{hyperref}
\newtheorem{theorem}{Theorem}

\newtheorem{lemma}[theorem]{Lemma}

\definecolor{Red}{cmyk}{0,1,1,0}

\definecolor{Blue}{cmyk}{1,1,0,0}

\newcommand{\mage}[1]{\langle \s_{#1}\rangle}
\newcommand{\ba}{\begin{array}}
\newcommand{\ea}{\end{array}}
\newcommand{\be}{\begin{equation}}
\newcommand{\ee}{\end{equation}}
\newcommand{\ben}{\begin{enumerate}}
\newcommand{\een}{\end{enumerate}}

\let\b=\beta

\let\e=\varepsilon

\let\g=\gamma

\let\o=\omega

\let\s=\sigma

\let\G=\Gamma
\let\L=\Lambda

\let\O=\Omega

\newcommand{\eop}{\nopagebreak\hfill\fbox{ }}

\newcommand{\R}{\mathbb{R}}

\newcommand{\Z}{\mathbb{Z}}

\begin{document}

\title{Phase
Transition in Ferromagnetic Ising Models with Non-Uniform External
Magnetic Fields}
\author{Rodrigo Bissacot\\
\footnotesize{Departamento Matem\'atica, UFMG, 30161-970 Belo Horizonte -- MG, Brasil}\\
\footnotesize{LMRS, Universit\'e de Rouen, F76801 Saint \'Etienne du Rouvray, France}\\
\footnotesize{\texttt{rodrigo.bissacot@gmail.com}} \and
Leandro Cioletti\\
\footnotesize{Departamento de Matem\'atica, Universidade de Brasília, 70910-900 Brasília -- DF, Brasil}\\
\footnotesize{\texttt{leandro.mat@gmail.com}}}

\maketitle

\begin{abstract}
In this article we study the phase transition phenomenon for the
Ising model under the action of a non-uniform external magnetic
field. We show that the Ising model on the hypercubic lattice with a
summable magnetic field has a first-order phase transition and, for
any positive (resp. negative) and bounded magnetic field, the model
does not present the phase transition phenomenon whenever $\liminf
h_i
> 0$, where ${\bf h} = (h_i)_{i \in \Z^d}$ is the external magnetic field.
\end{abstract}

\section{Introduction}
\label{intro}  The Lee-Yang theorem \cite{LY} is one of the most
revisited results in Statistical Mechanics \cite{Li-Ru,Li-So,Ne},
especially because of its application to the study of phase
transition phenomena. One consequence of this theorem is that for
any nonzero {\it uniform} magnetic field, i.e. ${\bf h}=(h_i)_{i \in
\Z^d}$, $h_i=h \in \mathbb{R}\backslash \{0\}$ for all $i \in \Z^d$
and $\beta = 1/kT$, the ferromagnetic $(J > 0)$ Ising model on
$\Z^d$
%$\Omega=\{-1,+1\}^{\Z^{d}}$($d\geq1$)
has an unique Gibbs measure in the thermodynamic limit, independently
of the boundary conditions.

In this paper we consider more general models where the magnetic
field ${\bf h}$ is not supposed to be uniform. For such models the
Lee-Yang Theorem is still valid, and a natural question is to ask if
for these models the Lee-Yang Theorem still implies the absence of
phase transition. The question of the uniqueness of the Gibbs
measure in a non-uniform positive magnetic field it was considered
by Georgii \cite{georgii-van} and Fontes and Neves \cite{LR}. They
considered the model with a non-negative random field of positive
mean, and proved that for all temperatures that there exists a
unique Gibbs state. Here, in the Theorem 4 we prove uniqueness of the
Gibbs states at all temperatures for all models for which $\liminf h_i> 0$ ( no average assumption is required for the fields $h_i$),   
showing that the Lee-Yang Theorem, in this case, still implies the absence of phase
transition. Although we present a proof for Theorem 4 in the
hypercubic lattice, the same argument works for any connected
amenable quasi-transitive graph with bounded degree.

It is well known that the Ising model with postive magnetic field can present phase transition depending on the graph structure of the model. Jonasson and Steif in \cite{JE} showed that the Lee-Yang Theorem does not imply the absence of a first-order phase transition for this model in any nonamenable graph. Basuev in \cite{Basu} obtained the same result for a class of amenable but not quasi-transitive graphs. In this paper, assuming that the magnetic field decays to zero, we obtain the same result for hypercubic lattices, which are examples of amenable and quasi-transitive graphs. In other words, even with the magnetic field taking positive values at all sites of the hypercubic lattice, we prove that the Ising model can present a first-order phase transition.

%
%%%%%%%%%%%%%%%%%%%%%%%%%%%%%%%%%%%%%%%%%%%%%%%%%%%%%%%%%%%%%%%%%%%%%%%%%%%%%%%%%%%%%%%%%%%%%%%%%%%%%%%%%%%%%%%
\section{Preliminaries and Main Results}
%%%%%%%%%%%%%%%%%%%%%%%%%%%%%%%%%%%%%%%%%%%%%%%%%%%%%%%%%%%%%%%%%%%%%%%%%%%%%%%%%%%%%%%%%%%%%%%%%%%%%%%%%%%%%%%

Consider the distance between $x$ and $y$ on $\Z^{d}$ given by
$\|x-y\|=\sum_{i=1}^{d}|x_i - y_i|$ and, for any finite $\Lambda
\subset \mathbb{Z}^d$, denote by $\partial \L$ the set of sites in
$\Z^d$ whose distance to $\L$ is equal to $1$. The energy in
$\Lambda$ of each configuration $\sigma \in
\O\equiv\{-1,+1\}^{\Z^{d}}$ satisfying the {\it boundary condition}
$\o \in \O$ in $\Lambda$ ($\o$ b.c.), that is, $\s_i=\o_i\ \forall\
i \in \Z^{d}\backslash\Lambda$, is given by the Hamiltonian
\begin{eqnarray}
\label{hamilton} H^{\o}_{\L}(\s)=\displaystyle-
\frac{J}{2}\sum_{\langle i,j\rangle}\s_i\s_j - \sum_{i\in\L}h_i\s_i,
\end{eqnarray}
where $\langle i,j\rangle$ denotes the set of ordered pairs in
$\L\cup\partial\L$ of nearest neighbors. For $J>0$ and ${\bf
h}=(h_i)_{i \in \Z^d}$ the Hamiltonian defines a ferromagnetic Ising
model with external field ${\bf h}$.

The {\it Gibbs measure in $\Lambda$ with $\o$ b.c.} is the
 probability measure on $(\O,\mathcal{F})$, where $\mathcal{F}$ is
 the sigma-algebra generated by the cylinder sets given by
\begin{eqnarray}
\label{gibbs_finita} \mu^{\b,{\bf h},\o}_{\Lambda}(\s)= \frac{e^{-\b
H^{\o}_{\L}(\s)}}{Z^{\o}_{\Lambda}}
\end{eqnarray}
if $\s$ satisfies the $\o$ b.c. and zero otherwise. The
normalization factor is the standard {\it partition function}
\begin{eqnarray}
Z^{\o}_{\Lambda}=\sum_{\s}e^{-\b H^{\o}_{\L}(\s)}
\end{eqnarray}
with the sum over all configurations $\s$ satisfying the $\o$ b.c.

We denote by $\mathcal{G}_{\b, {\bf h}}$ the set of {\it Gibbs
measures} given by the closed convex hull of the set of weak limits:
\begin{equation}
\mu^{\b,{\bf h},\o} = w-\lim_{\L_n\nearrow\Z^d}\mu^{\b,{\bf
h},\o}_{\Lambda_n}
\end{equation}
where $\L_n \subset \L_{n+1}$ and $\o$ runs over all boundary
conditions.

We say that there is a first-order {\it phase transition} when the
set $\mathcal{G}_{\b, {\bf h}}$ contains more than one measure, see
\cite{E,G}. In the case of a non-zero uniform field, the Lee-Yang
theorem can be used to prove that the analyticity of the pressure
with respect to the parameter $h$ is equivalent to saying that the
system has no phase transition, see \cite{Ru}.

We write ${\bf h}\in \ell^1(\Z^d)$ when $\|{\bf h}\|_1 =
\sum_{i\in\Z^d}|h_i|<\infty$ and ${\bf h} \in \ell^{\infty}(\Z^d)$
when ${\bf h}$ is bounded, that is, $\sup_{k \in \Z^d} |h_k| <
\infty $.

The paper is organized as follows: Our main result is presented in
Section 3, roughly speaking it states that, if the external field
${\bf h}$ is summable on $\Z^d$, that is, ${\bf h}\in \ell^1(\Z^d)$,
then the model has a first-order phase transition. This result is
obtained by a Peierls-type argument, based on contours already used
in \cite{Pf,Ve} and a cancelation argument for clusters of contours
with opposite signs.

In Section 4, we adress the question of non-summable {\bf h}. We
show that, if we have $\liminf h_i > 0$ (similarly for $\limsup h_i
< 0$), then the system has no first-order phase transition. We
remark that although this fact is pretty obvious, from the physical
point of view, the arguments we found in the literature are based
upon translation-invariance \cite{L2}. For a general
$\ell^{\infty}(\Z^d)$ magnetic field it is not clear that we can use
those techniques. By controlling the Radon-Nikodyn derivative we
show how to deal with the non-invariant case.

\section{The expansion in contours and the phase transition}

The arguments in this section can be generalized for higher
dimensions, we take dimension $2$ for simplicity. We will describe
roughly the {\it contours} in $\Z^2$, for details the readers can
see \cite{Pf,Ve}. This approach uses a bijection between finite
collections of contours in ${\Z^2}^{*}= \Z^2 + (1/2,1/2)$ and
configurations with some fixed boundary  condition. Without loss of
generality, we always suppose that the finite set $\L$ is a square.
This fact helps us to assure that, if we have a finite number of
contours in the dual set of $\L$ and already fixed a boundary
condition, then there is a configuration associated to these
contours,
see lemma 2.1.2 \cite{Ve}.\\

\begin{theorem}
If the magnetic field {\bf h} in the hamiltonian in (\ref{hamilton})
belongs to $\ell_1(\Z^2)$, then the model presents a phase
transition when $J>3\|{\bf h}\|_1$.
\end{theorem}
{\it Proof:} let $\L\subset\Z^2$ be a finite set and suppose
initially that $h_i\geq 0\ \forall \ i \in \Z^2$. It follows from
the second Griffiths inequality that
$$
\begin{array}{rcl}
\langle\s_i\rangle^{\b,{\bf h},+}_{\L}&\geq&
\langle\s_i\rangle^{\b,{\bf 0},+}_{\L} = 1-2\mu^{\b,{\bf
0},+}_{\L}(\{\s_i=-1\}),
\end{array}
$$
where ${\bf 0}$ is the null magnetic field. Using this lower bound
and the following standard inequality (see \cite{OV} page 170 for
details)
$$ \mu^{\b,{\bf
0},+}_{\L}(\{\s_i=-1\})\leq
c(\b):=\displaystyle\sum_{n=4}^{\infty}(2n+3)3^{n-1}e^{-2\b Jn},
$$
since clearly $c(\b)\to 0$, uniformly in $\L$ when $\b\to\infty$, we
get that \be \label{eq:exp-val-to-one}
\displaystyle\lim_{\b\to\infty}\langle\s_i\rangle^{\b,{\bf h},+} =
1. \ee

  The next step is to show that
$\lim_{\b\to\infty}\langle\s_i\rangle^{\b,{\bf h},-}=-1$. For this,
it will be convenient to consider the  Hamiltonian of the Ising
model with -- b.c. given by
 \be \label{ham-ising}
H^{-}_{\L}(\s)=\displaystyle-\sum_{\langle
i,j\rangle}\frac{J}{2}(\s_i\s_j-1) - \sum_{i\in\L}h_i(\s_i+1). \ee
Note that this normalization does not change the measure
$\mu^{\b,{\bf h},-}_{\L}$.

We will identify $\Z^2$ with the subset of $\mathbb{R}^2$ of integer
coordinates. Fix a finite set $\L \subset \Z^2$, to each site $i$ of
$\L$ we associate the dual plaquette $p^{*}(i)$ having $i$ at its
center. We call {\it plaquettes} the unit squares in $\mathbb{R}^2$
whose corners are in ${\Z^2}^{*}$. The {\it dual set} $\L^{*}$ of
$\L$ is the subset of ${\Z^2}^{*}$ of the corners of $p^{*}(i)$,
where $i$ is some site of $\L$.

For each configuration $\s_\L$ satisfying the -- b.c. we will
associate a family of contours as follows: for each pair of nearest
neighbors sites in $\s_\L$, where we have opposite signs, we
consider the unit segment $e$ joining the two sites. The dual
segment $e^*$ will be the unit segment orthogonal to $e$ passing
through the middle point of $e$, and joining the two sites of
$\L^{*}$ that are closest to that middle point.

The union of these dual unit segments, called {\it edges}, will form
closed curves in $\mathbb{R}^2$ such that the sites in $\L^{*}$ will
have degree $0,2$ or $4$ and, using some rule we can
\textquotedblleft cut\textquotedblright the corner when the degree
is 4, see \cite{Ve} for details.

This process give us a bijection between the configurations
satisfying -- b.c. and the finite sets of {\it compatible} contours
in $\L^{*}$, that is, closed self-avoiding and pairwise mutually
avoiding contours in $\L^{*}$.

Fixed the -- b.c., for each contour $\g$ there is an unique
configuration $\s_{\g}$ which has $\g$ as unique contour. We define
the {\it interior of} $\g$, $\mbox{int}\g$, as the set of all
$i\in\Z^2$ such that, for $\s_{\g}$, we have $\s_i=+1$ and {\it
$d(i,\g)>1$}, where $d$ denote the Euclidean distance in
$\mathbb{R}^2$. We will use the notation $\overline{\ \mbox{int}\g}$
for the set of all $i\in\Z^2$ for which $\s_i=+1$ in $\s_{\g}$. The
volume of $\g$ is the cardinality of $\overline{\ \mbox{int}\g}$,
$\mbox{vol}\g=|\overline{\ \mbox{int}\g}|$.

Let $\s_{\L}$ be a configuration satisfying the -- b.c.. We have
already seen that there is a finite set of contours associated to
$\s_{\L}$. For each contour $\g$ all spins at
$\overline{\mbox{int}\g}\setminus\mbox{int}\g$ have the same value.
We say that $\g$ is of type + (resp. type --), if the value of these
spins are $+1$ ( resp. $-1$).

We define over the set of signed contours a function $\xi$ given by
\be \xi(\g)= \left\{
\begin{array}{rl}
\exp{\left(-2\b J|\g|\ - \ 2\b\!\!\textstyle \sum\limits_{ i\in
\overline{\mbox{\ {\scriptsize int}} \g} }h_i\right)},&\ \g \
\mbox{of type}\ -;\\[1cm]
\exp{\left(-2\b J|\g|\ + \ 2\b\!\!\textstyle\sum\limits_{ i\in
\overline{\mbox{\ {\scriptsize int}} \g} }h_i \right)},&\ \g \
\mbox{of type}\ +,
\end{array}
\right. \ee where $|\g|$ denote the number of unit segments that
compose the contour.

Denoting by $Z^{-}_{\L}$ the partition function corresponding to the
-- b.c., it follows from a straightfoward computation that:
\begin{eqnarray}
Z^{-}_{\L}=1+\displaystyle\sum_{n\geq 1}\displaystyle\frac{1}{n!}
\sum_{\substack{(\g_1,\ldots,\g_n)\\[0.05cm] \L^* - {\scriptsize\mbox{compatibles}}}}\prod_{k=1}^n\xi(\g_k).
\end{eqnarray}

We remark that the value $\xi(\g)$ depends explicitly on the type of
the contour, which is different from the usual cluster expansion.
Since
\begin{eqnarray}
\langle\s_i\rangle^{\b,{\bf h},-}_{\L} =2\ \mu^{\b,{\bf
h},-}_{\L}(\{\s_i=+1\})-1
\end{eqnarray}
if we prove that $\mu^{\b,{\bf h},-}_{\L}(\{\s_i=+1\})\to 0$,
uniformly in $\L$ when $\b\to\infty$, we are done.

 Whenever $\s_i=+1$ there
exists a contour type +, denoted by $\stackrel{+}{\g}$, that
involves the site $i$. We use the notation $\stackrel{+}{\g}\odot \
i$ to indicate that $\stackrel{+}{\g}$ involves the site $i$.
 We say that $\g$ involves $\g'$, and we write $\g\odot\g'$, if any site $i\in\Z^d$ involved by $\g'$, it is also involved by $\g$.

Before continuing the theorem's proof we need the following lemma:

\begin{lemma}
 Let $\{\g_1,\ldots,\g_n\}$ be a collection of $\L^*$-compatible signed contours. If ${\bf h}\in\ell^1(\Z^d)$, then
$$
e^{-2\b\|h\|_1}\left(\prod_{k=1}^n e^{-2\b
J|\g_k|}\right)\leq\prod_{k=1}^n \xi(\g_k)\leq
e^{2\b\|h\|_1}\left(\prod_{k=1}^n e^{-2\b J|\g_k|}\right).
$$
\end{lemma}
\noindent{\it Proof:} For a fixed collection of $\L^*$-compatible
signed contours $\{\g_1,\ldots,\g_n\}$ the relation involving
$\odot$ determines naturally a partial
 order in this set. Let $\{\g_{r_1},\ldots, \g_{r_k}\}$ be the set of all maximal elements with respect to this partial order.
 So we have that
\be \label{eq:canc1} \prod_{k=1}^n \xi(\g_k)= \prod_{l=1}^k
\xi(\g_{r_l})\left( \prod_{\{j:\g_{r_l}\odot\g_j\}}\xi(\g_j)
\right), \ee where the product over an empty set is equal to one. If
$\{j:\g_{r_l}\odot\g_j\}=\emptyset$ for all $1\leq l\leq k$, the
upper and lower bounds claimed in the lemma are straightforward
since $\overline{\mbox{int}\g_{r_l}}\cap
\overline{\mbox{int}\g_{r_m}}=\emptyset$ whenever $1\leq m< l\leq
k$. On the other hand, for each $1\leq l\leq k$ such that the set
$\{j:\g_{r_l}\odot\g_j\}$ is not empty, consider the graph
$G_l=(V_l,E_l)$, where the vertex set $V_l=\{\g_{r_l}\}\cup\{\g_j:
\g_{r_l}\odot\g_j\}$ and the edges set $E_l$ is the collection of
all unordered pairs $\{\g_j,\g_l\}$ such that if $\g_m\odot\g_j$ or
$\g_m\odot\g_l$, then $\g_m\odot\g_j$ and  $\g_m\odot \g_l$, i.e.,
there are no intermediate contours between $\g_j$ and $\g_l$.

From the definition of the contours and the edges set it is easy to
see that $G_l$ is a rooted tree where the root is the  most exterior
contour, i.e. this contour is not in the interior of any other
contour. For $\g_j,\g_s\in V_l$, we denote by $d(\g_j,\g_s)$ the
usual graph distance in $G_l$. Let $TV_l$ be the set of vertices in
$V_l$ that are in the last generation of the tree, i.e., the
vertices with degree one on $G_l$ and that the distance to the root
is maximal. Consider the following partition of
$$
TV_l= \displaystyle\bigcup_{j=1}^{t(l)} TV_l(j),
$$
where $t(l)$ is the cardinality of the set $\{\g\in
V_l:d(\g,TV_l)=1\}$  and for each $1\leq j\leq t(l)$,
 the sets $TV_l(j)=\{\g_{j1},\ldots,\g_{jr_j(l)} \}$ are the maximal
  subsets of $TV_l$ possesing the same parent $\g_{P_l(j)}$. By rearranging, the product (\ref{eq:canc1}) can be written as
\be \label{eq:canc2}
\prod_{l=1}^{k}\left[\prod_{j=1}^{t(l)}\left(\xi(\g_{P_j(l)})
\prod_{\g\in TV_l(j)}\xi(\g)\right)\prod_{\g\in
\tilde{G}_l}\xi(\g)\right], \ee where $\tilde{G}_l$ is the tree
subgraph of $G_l$ induced by the vertices
$$V_l\backslash (TV_l\cup \{\g_{P_1(l)},\ldots,\g_{P_{t(l)}(l)}\}).$$

Observe that the value of the function $\xi$ at the root $\g_{r_l}$
appears in the product $\prod_{\g\in \tilde{G}_l}\xi(\g)$ when the
graph $\tilde{G}_l$ is not empty. Putting
$$
S_l(j)=\overline{\mbox{int}\g_{P_j(l)}}\ \backslash\bigcup_{\g\in
TV_j(l)}\overline{\mbox{int}\g},
$$
it follows from the definition of $\xi$, independent of the sign of
the contour $\g_{P_j(l)}$, that the following bounds hold:
$$
\displaystyle e^{-2\b\sum_{i\in
S_l(j)}h_i}\left(\prod_{\{\g_k:\g_k\in TV_l(j)\cup\{\g_{P_j(l)}\}\}}
e^{-2\b J|\g_k|}\right)\leq\xi(\g_{P_j(l)})\prod_{\g\in
TV_l(j)}\xi(\g),
$$
and
$$
\xi(\g_{P_j(l)})\prod_{\g\in TV_l(j)}\xi(\g)\leq\displaystyle
e^{+2\b\sum_{i\in S_l(j)}h_i}\left(\prod_{\{\g_k:\g_k\in
TV_l(j)\cup\{\g_{P_j(l)}\}\}} e^{-2\b J|\g_k|}\right).
$$

Proceeding as above, for each $\tilde{G}_l$, we get new $S_l(j)$'s
that are disjoint for the ones previously defined. So using the
reasoning iteratively we finish the proof. The proof ends when each
of $\tilde{G}_l$ is only the root $\g_{r_l}$ or the root $\g_{r_l}$
and the elements connected to it.

\eop

To finish the proof of the theorem we use the lemma above and, by
the contour representation, we get the following upper bounds for
$\mu^{\b,{\bf h},-}_{\L}(\{\s_i=+1\})$:
\begin{eqnarray} \label{eq:mumenos}
&&\hspace{-0.5cm}\mu^{\b,{\bf h},-}_{\L}(\{\s_i=+1\}) \leq
\mu^{\b,{\bf
h},-}_{\L}(\{\exists\ \stackrel{+}{\g}\odot i\}\big)\nonumber\\[0.4cm]
&\leq& \displaystyle\frac{\sum_{\stackrel{+}{\g}\odot
i}\xi(\stackrel{+}{\g})
 \left( 1+ \sum_{n\geq1}\sum_{\substack{\{\g_1,\ldots,\g_n\}\cap\{\stackrel{+}{\g}\}=\emptyset\\
\L^*-{\scriptsize\mbox{compatibles}}}} \ \ \
\prod_{k=1}^n\xi(\g_k)\right)}
{1 + \sum_{n\geq 1}\sum_{\substack{\{\g_1,\ldots,\g_n\}\\
\L^*-{\scriptsize\mbox{compatibles}}}} \prod_{k=1}^n\xi(\g_k)}\nonumber\\
%%%%%%% inicio modificação 1 %%%%%%%%
%%%%%%%%%%%Fim modificação 1%%%%%%%%%%%%%
%%%%%%% inicio modificação 2 %%%%%%%%
&\leq& \displaystyle\frac{\sum_{\stackrel{+}{\g}\odot
i}\xi(\stackrel{+}{\g})
 \left( e^{2\b\|{\bf h}\|_1}+ \sum_{n\geq1}\sum_{\substack{\{\g_1,\ldots,\g_n\}\cap\{\stackrel{+}{\g}\}=\emptyset\\
\L^*-{\scriptsize\mbox{compatibles}}}} \ \ \ e^{2\b\|{\bf
h}\|_1}\left(\prod_{k=1}^n e^{-2\b J|\g_k|}\right)\right)}
{e^{-2\b\|{\bf h}\|_1} + \sum_{n\geq 1}\sum_{\substack{\{\g_1,\ldots,\g_n\}\\
\L^*-{\scriptsize\mbox{compatibles}}}} e^{-2\b\|{\bf
h}\|_1}\left(\prod_{k=1}^n e^{-2\b J|\g_k|}\right)}\nonumber
\\
%%%%%%%%%%%Fim modificação 2%%%%%%%%%%%%%
%%%%%%% inicio modificação 3 %%%%%%%%
%%%%%%%%%%%Fim modificação 3%%%%%%%%%%%%%
&\leq& \displaystyle\sum_{\stackrel{+}{\g}\odot
i}\xi(\stackrel{+}{\g})e^{4\b\|{\bf h}\|_1} \leq \sum_{{\g}\odot i}
\exp\left(-2\b J|\g|\ + \ 2\b\!\!\textstyle\sum\limits_{ i\in
\overline{\mbox{\ {\scriptsize int}} \g} }h_i+4\b \|{\bf
h}\|_1\right).
\end{eqnarray}

Using the last inequality and $|\g|\geq 4$, we obtain
\begin{eqnarray}
\mu^{\b,{\bf h},-}_{\L}(\{\s_i=+1\})\leq \sum_{n\geq
4}\exp\left(-2\b (Jn-3\|{\bf h}\|_1)\right)n3^n.
\end{eqnarray}

Taking  $J$ such that  $J>3\|{\bf h}\|_1$, we conclude that
$\mu^{\b,{\bf h},-}_{\L}(\{\s_i=+1\})\to 0$ when $\b\to\infty$,
uniformly in $\L$.

The case ${\bf h}\leq 0 \ (h_i\leq 0, \forall \ i \in \Z^2)$ with
{\bf h} $\in \ell_1(\Z^2)$ follows from the first case and the FKG
inequality. For an arbitrary field ${\bf h}$ in $\in \ell_1(\Z^2)$
we use FKG and reduce the problem to one of the previous cases. \qed
\\
\\
{\bf Corollary.} {\it For all the magnetic field \ ${\bf h}\in \ell_1(\Z^d)$,  the Ising model presents a firts-order phase transition.
}
\\

{\it Proof:} First replace the fields $h_i$ by zero, at all sites of a
finite large enough region $\G$, in order to have $J>3\|{\bf
\tilde{h}}\|_1$, where ${\bf\tilde{h}}$ denotes the modified magnetic
field. Thus there are two different measures $\mu_{-} \ \mbox{and}\
\mu_+$, by Theorem 1. Now, take the local function $A(\s)= \sum_{i
\in \G} h_i \sigma_i$. For any local function $f:\O\to\R$ we have that 
\begin{eqnarray*}%\label{medida}
\mu_-(fe^{\b A}):=\int_{\O} f(\s) e^{\b A(\s)} d\mu_-=\lim_{\L\nearrow\Z^d}\frac{\langle f\rangle^{\b,{\bf h},-}_{\L}}{\langle e^{-\b \sum_{i \in \G} h_i \sigma_i}\rangle^{\b,{\bf h},-}_{\L}}=
\frac{\langle f\rangle^{\b,{\bf h},-}}{\langle e^{-\b \sum_{i \in \G} h_i \sigma_i}\rangle^{\b,{\bf h},-}}.
\end{eqnarray*}
Analogously for the $+$ boundary condition.

Suppose by absurd that the model with the magnetic field ${\bf h}$
does not present a first-order phase transition, thus for any $\b>0$ fixed
and for all local function $f$ we have $\langle f\rangle^{\b,{\bf
h},-}=\langle f\rangle^{\b,{\bf h},+}$. 
Applying this equality for $f(\s)=e^{-\b A(\s)}$, we get 
$$\langle e^{-\b \sum_{i \in \G} h_i \sigma_i}\rangle^{\b,{\bf
h},-}=\langle e^{-\b \sum_{i \in \G} h_i \sigma_i}\rangle^{\b,{\bf
h},+}.$$

So it follows from the above equalities that  $\mu_{-} (f e^A)=\mu_{+} (f e^A)$ for
any $\b>0$ and for any local function $f$, which is in contradiction with Theorem 1 by taking $\b>0$ sufficiently large and
$f(\s)=\s_ie^{-A(\s)}$.  \qed \\

\noindent{\bf Remark:} In the above argument we show how we can work with
contours in a non-symmetric set up and also illustrate that a
finite-energy, (quasi-)local change in a infinite system does not
change global properties, such as (non)uniqueness of the Gibbs
state. In fact, the summability guarantees that this proof works
but, it is not a necessary condition for the phenomenon. For the
Semi-Infinite Ising Model on $\Z\times\Z_+$ Fr\"ohlich and Pfister
\cite{pfister-van} showed that there are two Gibbs states at
sufficiently low temperature with a constant magnetic field $h$ only
over the sites in the boundary of the lattice and, as we already
mentioned before, Basuev \cite{Basu} obtained a phase transition in
a more interesting case, where the magnetic field is constant at all
sites of the lattice $\Z^2\times\Z_+$.

\section{Absence of Phase Transition}

In this section we show the absence of phase transition in the
non-summable case under condition that the magnetic field satisfies $\liminf h_i>0$ for positive fields and
$\limsup h_i<0$ for negative ones. Although the proofs of the Lemma 3 and Theorem 4 below are presented for the hypercubic lattice, they are the same for any quasi-transitive connected amenable graph with uniformly bounded
degree due the Theorem 5 of \cite{JE}. For the proof of the Theorem 4, in this section, we follow close \cite{LR}.
\begin{lemma}\label{lema}
There is no phase transition in ferromagnetic Ising models with
uniform nonzero external magnetic field outside of a finite volume.
\end{lemma}
{\it Proof:} Let {\bf h} the magnetic field and $\L_0
\subset \Z^d$ a finite set such that, $h_i = h$ for all $i \notin
\L_0$ and $h \in \R\backslash\{0\}$. To show absence of phase
transition for this model it is enough to show that $
\mage{i}^{\b,{\bf h},-}=\mage{i}^{\b,{\bf h},+} $ for all $\b>0$ and
$i\in\Z^d$.

Consider $\Lambda_0 =\{k\} \subseteq \Lambda\subset \Z^d$, from now
$\langle \ . \ \rangle ^{\b,h,\omega}_{\L}$ and
$Z^{\b,{h},\omega}_{\Lambda}$ denote respectively the expected value
and the partition function with respect to the Gibbs measure defined
by the Hamiltonian (\ref{hamilton}) with boundary condition $\omega$
and constant magnetic field $h$. It follows from the definition
that, for all $ i \in \Lambda$
\begin{eqnarray}\label{lala}
\mage{i}^{\b,{\bf h},\omega}_{\Lambda} = \displaystyle\langle
\s_i\cdot e^{\b (h_k-h)\s_k}
\rangle^{\b,h,\omega}_{\Lambda}\frac{Z^{\b,{h},\omega}_{\Lambda}}{Z^{\b,{\bf
h},\omega}_{\Lambda}}.
\end{eqnarray}

We know that the expected value $\langle \s_i\cdot e^{\b
(h_k-h)\s_k} \rangle^{\b,h,\omega}_{\Lambda}$ is independent of the
boundary conditions in the thermodynamical limit. So, we need to
show that $Z^{\b,{h},\omega}_{\Lambda}/ Z^{\b,{\bf
h},\omega}_{\Lambda}$ it is also independent of the boundary
conditions when $\Lambda \nearrow \Z^d$.

In order to evaluate the limit of the above ratio we will need to
consider boundary conditions in $\Lambda\backslash\{k\}$. We define
$\o_1$ as $(\o_1)_{i}=\o_i$ for all $i\in\partial\Lambda$ with
$(\o_1)_{k}=+1$ and, $\o_2$ by $(\o_2)_i=\o_i$ for all
$i\in\partial\Lambda$ and $(\o_2)_{k}=-1$ so
\begin{eqnarray}
\displaystyle\frac{Z^{\b,{h},\omega}_{\Lambda}}{Z^{\b,{\bf
h},\omega}_{\Lambda}} = \displaystyle\frac{ e^{\b h} \cdot
Z^{\b,{h},\omega_1}_{\Lambda\backslash\{k\}} + e^{-\b h} \cdot
Z^{\b,{h},\omega_2}_{\Lambda\backslash\{k\}}        } {e^{\b h_k}
\cdot Z^{\b,{h},\omega_1}_{\Lambda\backslash\{k\}} + e^{-\b h_k}
\cdot Z^{\b,{h},\omega_2}_{\Lambda\backslash\{k\}}}
\end{eqnarray}

To show that the above expression in the thermodynamic limit is
independent of the boundary condition $\o$, it is enough to show
that
$Z^{\b,{h},\omega_2}_{\Lambda\backslash\{k\}}/Z^{\b,h,\omega_1}_{\Lambda\backslash\{k\}}$
does not depend on $\o_1$ and $\o_2$. To do this we take $i = k$ in
(\ref{lala}) and $h_k \rightarrow +\infty$, and we obtain

\begin{eqnarray*}
1 &=& \displaystyle
\mu^{\b,h,\omega}_{\Lambda}(\{\s_k=+1\})\lim_{h_k\to\infty}\left(
e^{\b (h_k-h)} \frac{ e^{\b h} \cdot
Z^{\b,{h},\omega_1}_{\Lambda\backslash\{k\}} + e^{-\b h} \cdot
Z^{\b,{h},\omega_2}_{\Lambda\backslash\{k\}}        }
{e^{\b h_k} \cdot Z^{\b,{h},\omega_1}_{\Lambda\backslash\{k\}} + e^{-\b h_k} \cdot Z^{\b,{h},\omega_2}_{\Lambda\backslash\{k\}}} \right)\\
&=&\mu^{\b,h,\omega}_{\Lambda}(\{\s_k=+1\})\left( 1 + e^{-2\b h}
\cdot
\frac{Z^{\b,{h},\omega_2}_{\Lambda\backslash\{k\}}}{Z^{\b,{h},\omega_1}_{\Lambda\backslash\{k\}}}\right)
\end{eqnarray*}

for all finite set $\Lambda$ containing $k$. Then,
\begin{eqnarray}
\lim_{\L\nearrow\Z^d}\frac{Z^{\b,{h},\omega_2}_{\Lambda\backslash\{k\}}}{Z^{\b,{h},\omega_1}_{\Lambda\backslash\{k\}}}
=e^{2\b h}\frac{\mu^{\b,h,\omega}(\{\s_k=-1\})
}{\mu^{\b,h,\omega}(\{\s_k=+1\}) }
\end{eqnarray}
and the argument follows by induction in $|\Lambda_0|$. \qed

\begin{theorem}
If ${\bf h} \in \ell_{\infty}(\Z^d)$ is such that $\liminf_{i \in
\Z^d} h_i
> 0$, then the Ising model with external field ${\bf h}$ has no
phase transition.
\end{theorem}
{\it Proof:} We omit the parameter $\beta$ since the
argument is valid for all $\beta > 0$. Let be $\overline{h} =
\sup_{i \in \Z^d}h_i$, $\underline{h}= \liminf_{i \in \Z^d}h_i$ and
$\e$ a positive number such that $0<\e<\underline{h}/2$.
We denote by $\L\subset\Z^d$ a finite subset containing $\{i\in \Z^d; |h_i|<\e\}$,
$h_{\Lambda}$ the restriction of the external field ${\bf h}$ to the volume $\L$,
$\overline{h}_{\Lambda}$ the constant magnetic field in the set
$\Lambda$ with the value $\overline{h}$, analogously for
$\underline{h}$ and $\underline{h}^{\e}_{\L}$ the constant magnetic field taking values $\underline{h}-\e$. It follows from Fundamental Theorem of Calculus
that for any finite $\Gamma\subset\Z^d$ with $\Lambda\subset\Gamma$,
we have that the difference
$$\mage{i}_{\Gamma}^{h_\L,\overline{h}_{\Gamma\backslash\Lambda},+} -
\mage{i}_{\Gamma}^{h_\L,\underline{h}^{\e}_{\Gamma\backslash\Lambda},-}$$
is equal to
\begin{eqnarray*}
\b\int_{\underline{h}-{\e}}^{\overline{h}}\sum_{j \in
\Gamma\backslash\Lambda }\langle \sigma_i;\sigma_j
\rangle^{h_\Lambda,x_{\Gamma\backslash\Lambda},+}_{\Gamma} \ dx +
\mage{i}_{\Gamma}^{h_\L,\underline{h}^{\e}_{\Gamma\backslash\Lambda},+} -
\mage{i}_{\Gamma}^{h_\L,\underline{h}^{\e}_{\Gamma\backslash\Lambda},-}.
\end{eqnarray*}

 By lemma \ref{lema} we know that there is no phase transition for the
ferromagnetic models with uniform nonzero magnetic field outside of
a finite volume. Then, by the FKG inequality taking the limit when
$\Gamma \nearrow \Z^d$ we have:

$$
0 \leq \mage{i}^{h_\L,\overline{h}_{Z^d\backslash\Lambda},+} -
\mage{i}^{h_\L,\underline{h}^{\e}_{Z^d\backslash\Lambda},-} =
\displaystyle
\lim_{\Gamma \nearrow\Z^d}\b\int_{\underline{h}-{\e}}^{\overline{h}}\sum_{j \in
\Gamma\backslash\Lambda}\langle \sigma_i;\sigma_j
\rangle^{h_\Lambda,x_{\Gamma\backslash\Lambda},+}_{\Gamma} \ dx
$$

Now, taking the limit when $\Lambda \nearrow \Z^d$ we obtain:

$$
0 \leq \mage{i}^{{\bf h},+} - \mage{i}^{{\bf h},-} \leq
\displaystyle\lim_{\Lambda \nearrow \Z^d} \displaystyle\lim_{\Gamma
\nearrow \Z^d}\b\int_{\underline{h}-\e}^{\overline{h}}\sum_{j \in
\Gamma\backslash\Lambda}\langle \sigma_i;\sigma_j
\rangle^{h_\Lambda,x_{\Gamma\backslash\Lambda},+}_{\Gamma} \ dx
$$

Since the truncated correlation functions are non-increasing in $h_k
( k \in \Z^d)$, we get
\begin{eqnarray*}
 0 \leq \mage{i}^{{\bf h},+} - \mage{i}^{{\bf h},-} &\leq&
\displaystyle\lim_{\Lambda \nearrow \Z^d} \displaystyle\lim_{\Gamma
\nearrow \Z^d}\b\int_{\underline{h}-\e}^{\overline{h}}\sum_{j \in
\Gamma\backslash\Lambda}\langle \sigma_i;\sigma_j
\rangle^{\underline{h}^{\e}_{\Lambda},x_{\Gamma\backslash\Lambda},+}_{\Gamma}
\ dx\\
 &=& \displaystyle\lim_{\Lambda \nearrow \Z^d}
\displaystyle\lim_{\Gamma \nearrow
\Z^d}\mage{i}^{\underline{h}^{\e}_{\Lambda},\overline{h}_{\Gamma\backslash\Lambda},+}_{\Gamma}-
\mage{i}^{\underline{h}^{\e}_{\Lambda},\underline{h}^{\e}_{\Gamma\backslash\Lambda},+}_{\Gamma}\\
&\leq& \displaystyle\lim_{\Lambda \nearrow \Z^d}
\mage{i}^{\underline{h}^{\e}_{\Lambda},+}_{\Lambda}-
\mage{i}^{\underline{h}-\e,+}\\
 &=& 0
\end{eqnarray*}

The last inequality comes again from FKG. By standard arguments
\cite{L1}, there is only one Gibbs state for the model. The
analogous result holds when the magnetic field is negative. \eop\\

\noindent{\bf Acknowledgments}\\

It is a pleasure to thank Aernout Van Enter for many
valuable comments and references. The authors also thank to Roberto
Fern\'{a}ndez for several useful references and thanks to Luiz
Renato Fontes, Arnaud Le Ny and specially Sacha Friedli, who point
out to us an error in the preliminary version, for fruitful
discussions personally and by mail. Rodrigo Bissacot was supported
by Conselho Nacional de Desenvolvimento Cient\'{i}fico e
Tecnol\'ogico (CNPq) and Leandro Cioletti was supported by Fundação
de Empreendimentos Científicos e Tecnológicos(Finatec).

\end{document}